\documentclass[twocolumn]{IEEEtrans}

\def\BibTeX{{\rm B\kern-.05em{\sc i\kern-.025em b}\kern-.08em
    T\kern-.1667em\lower.7ex\hbox{E}\kern-.125emX}}

\begin{document}

\title{Bayesian Estimation for Land Surface Temperature Retrieval:  The Nuisance of Emissivities}

\author{J. A. Morgan\\The Aerospace Corporation\\P. O. Box 92957\\Los Angeles, CA 90009}



\maketitle

\begin{abstract}
An approach to the remote sensing of land surface temperature is developed using the methods of 
Bayesian inference.  The starting point is the maximum entropy estimate for the posterior distribution 
of radiance in multiple bands.  In order to convert this quantity to an estimator for surface 
temperature and emissivity with Bayes' theorem, it is necessary to obtain the joint prior probability 
for surface temperature and emissivity, given available prior knowledge.  The requirement that any 
pair of distinct observers be able to relate their descriptions of radiance under arbitrary Lorentz 
transformations uniquely determines the prior probability.  Perhaps surprisingly, surface temperature 
acts as a scale parameter, while emissivity acts as a location parameter, giving the prior probability
\[
P(T,\epsilon\mid K)=\frac{const}{T}dTd\epsilon 
\]
Given this result, it is a simple matter to construct estimators for surface temperature and 
emissivity.  A Monte Carlo simulation of land surface temperature retrieval in selected MODIS bands 
is presented as an example of the utility of the approach.
\end{abstract}

\begin{keywords}
Remote Sensing, Land Surface Temperature, Sea Surface Temperature.
\end{keywords}

\section{Introduction}
\PARstart{T}{his} paper derives the joint prior probability for surface temperature and emissivity for 
the land surface temperature retrieval problem in remote sensing.  It presents analysis necessary for 
formulating a Bayesian approach to that problem, together with a Monte Carlo simulation of land 
surface temperature (LST) and surface emissivity extractions.

After a brief description of the problem and the method of attack, the maximum entropy estimator for 
the mismatch between sensed and forward model radiance is given.  Next, the joint prior probability 
for surface temperature and emissivity is obtained.  This quantity is required in order to construct 
a useable estimator for surface temperature and emissivity.  With the prior probability in hand, it 
is a simple matter to construct expressions for the expected values of surface temperature and 
emissivity consistent with sensor aperture radiances and available prior knowledge.  
Finally, a sample temperature-emissivity separation is presented using MODTRAN calculations both for 
the forward model and for simulated sensor radiances. 
\newline
\copyright 2004 The Aerospace Corporation

\section{The Temperature-Emissivity Separation Problem amd its Discontents}
Increasingly capable remote sensing technology has sparked interest in exploiting thermal 
emission from surfaces both for remote sensing of surface temperature and of emissivity.  On the one 
hand, surface temperature studies form a significant portion of the science goals of MODIS, ASTER, 
and MTI, while AVHRR has been used operationally for sea surface and land surface temperature studies 
for many years.  On the other, the use of emissivity information in thermal portions of the spectrum 
for geological remote sensing has grown rapidly in recent years, and is as prominent 
in the science goals of MODIS and similar instruments as is surface temperature.  

Accordingly, the problem of temperature-emissivity separation merits close examination.  The entirety 
of the information about a radiating surface comes from the thermal radiation it emits, 
conventionally parameterized as the product of blackbody radiation at the surface temperature $T$ and 
the emissivity $\epsilon_{k}$,
\begin{equation}
I_{s}=\epsilon_{k}B_{k}(T) \label{eqn1}
\end{equation}
at each wavenumber $k$.  Suppose one observes a radiating patch of a surface at each of $n$ wavenumber 
intervals, and that one knows how to correct for the effects of line-of-sight (LOS) attenuation and 
contamination by radiance from other sources.  One then has $n$ equations of the form (\ref{eqn1}), but 
$n+1$ unknowns, including the surface temperature.  In the absence of knowledge about $T$ or 
$\epsilon_{k}$ from extraneous sources, one has an underdetermined problem. 
	
A variety of methods has been proposed for handling the temperature-emissivity separation (TES) 
problem\cite{Dash2002}.  In most approaches to this problem, simultaneous LST and band emissivity retrieval 
depends
upon specifying an emissivity value in one or more reference bands.  The MODIS Land Surface Temperature
(LST) algorithm\cite{Wan1999} seeks a pair of reference channels in a part of the 
thermal spectrum in 
which the emissivity of natural surfaces displays very limited variation, and may therefore be regarded 
as known with good confidence.  Multiband emissvities inferred on this basis are called "relative" 
emissivities\cite{Li1999}.  Algorithms using this approach include the reference channel 
method\cite{KahleAlley1992} and emissivity normalization\cite{KealyHook1993}.  In the former, a value 
of emissivity is assumed for one band, and in the latter, an approximate surface temperature is 
obtained by noting that emissivity cannot exceed unity, in order to close the system of equations.  
Other relative emissivity retrieval methods include the temperature-independent spectral index 
method\cite{PetitcolinVermote2002},\cite{LiBecker1993} and spectral ratios\cite{Watson1992}.  
The study by Li et al.\cite{Li1999}
shows that all of these relative emissivity retrieval algorithms are closely related, and argues that 
they may be expected to show comparable performance.  A different approach has been proposed for 
analysis of Multispectral Thermal Imager (MTI) data\cite{BorelSzymanski1998}, in which radiances are 
collected from a surface with looks at nadir and 60 degrees off-nadir, assuming a known angular 
dependence of emissivity, in order to balance equations and unknowns.  The generalized split-window 
LST algorithm\cite{WanDozier1996} likewise uses dual looks in a regression-law based approach. 
The basis of the "grey body emissivity" 
approach\cite{BarducciPippi1996} is the slow variation of emissivity with wavelength for certain 
natural targets.  The physics-based MODIS LST 
algorithm\cite{WanLi1997} exploits observations taken at day and at night, on the assumption that band 
emissivites do not change over periods of a few weeks.  

It is clear that the methods described depend upon \emph{a priori} assumptions about the variation of 
emissivity, either with wavelength, or with look angle, or over time, from which one would like to be 
free.  The work described in this paper uses Bayesian inference to 
retrieve estimates of surface parameters.  This approach allows one to treat emissivities as "nuisance" 
parameters which may be integrated out of a posterior distribution function between parsimoniously 
chosen, and hence "uninformative," limits.

It might appear odd to use as an approach to the separation of temperature and 
emissivity a Bayesian estimator which, in essence, allows one to ignore the actual value of 
emissivity.   Equation (\ref{eqn1}) shows that thermal radiance is linear in emissivity.  However, 
the Planck 
function goes as a fairly high power of the temperature in the LWIR, and is close to exponential in 
temperature in the MWIR.  Any roughness in the treatment of sensed radiance-as in allowing the 
assumed emissivity in the estimator to take on a wide range of values-may, therefore, be expected to 
lead to comparatively small errors in the inferred surface temperature.  In fact, it turns out that 
the posterior distribution for surface temperature to be developed gives sharp limits to allowable 
surface temperature even in the presence of considerable latitude in the value of possible 
emissivities.  In most cases, only a narrow range of surface temperature is consistent with the 
sensed radiance in multiple bands, whatever be assumed about emissivity.  

Once a reasonably good estimate of surface 
temperature is in hand, it is a simple matter to insert it into estimators for the individual band 
emissivities, and for the uncertainty in those values consistent with available knowledge.  The 
\emph{a priori} limits on emissivity may then be contracted, and a new estimate of surface 
temperature obtained.  The expectation values of surface temperature and emissivity may thus be 
refined iteratively.  It is also possible to search for a (local) maximum for the posterior 
likelihood for these parameters.  Because the TES problem is underdetermined, this will not give a 
unique global maximum, but, given the insensitivity of surface temperature estimates to small 
emissivity errors, the local maximum may be expected to give results close to the physical values for 
the parameters of interest.
 
\section{Maximum Entropy Estimators for Surface Parameters}
Consider the problem of estimating surface temperature and emissivity from radiance detected by a 
remote sensor.  The sensor supplies measurements of radiance $I$ at the aperture.  A forward model is 
required to compute the value of aperture radiance as a function of, among other things, the 
parameters we wish to extract.  Assume initially (for simplicity) that the sensor has fine spectral 
resolution.  The forward model radiance may be described at each wavenumber $k$ by a form of the 
Duntley equation~\cite{Duntley1948}
\begin{equation}
I_{F}(k)=\epsilon_{k}B_{k}(T)exp(-\frac{\tau_{k}}{\mu})+
\frac{\rho_{k}}{\pi} F_{k}^{\downarrow}
(0)exp(-\frac{\tau_{k}}{\mu})+I_{k}^{\uparrow}(\tau,\mu) \label{eqn2}
\end{equation}
$I_{k}^{\uparrow}(\tau,\mu)$ and $F_{k}^{\downarrow}(0)$ are the upwelling diffuse radiance at nadir 
optical 
depth $\tau$ (top of the atmosphere, or TOA, for spaceborne sensors; $\mu$ is the cosine of the angle 
with respect to zenith) and the downwelling irradiance at the surface, respectively.  $B_{k}(T)$ is the 
Planck function at surface temperature $T$.  The emissivity is $\epsilon_{k}$, and the surface 
reflectance $\rho_{k}=1-\epsilon_{k}$.  The form of (\ref{eqn2}) is what one would get assuming a
Lambertian surface obeying Khirchoff's law.  The analysis presented below does not depend upon a 
Lambertian approximation to surface reflectances; in fact, it makes no assumption regarding their 
angular behavior~\cite{footnote1}.  In what follows it will be assumed that the only unknown 
quantities in the preceding equation are T and  $\epsilon_{k}$.  Generalization of 
the analysis which follows to the case of reflectance not equal to one minus the 
emissivity poses no difficulties.  

An estimator for the probability that, given observed radiances $I(k)$, the surface parameters $T$ 
and $\epsilon_{k}$ take on particular values, is constructed in the following way
\cite{Bretthorst1987},\cite{Bretthorst1988},\cite{Bretthorst1988a}, (\emph{vide.} also 
a related discussion in Landau and Lifschitz\cite{LandauLifschitz1958}, pp. 343-5).  The posterior 
probability for $T$ and $\epsilon_{k}$ is given by Bayes' theorem as
\begin{equation}
P(T,\epsilon_{k}\mid I,K)=
P(T,\epsilon_{k}\mid K)\frac{P(I\mid T,\epsilon_{k},K)}{P(I\mid K)} \label{eqn3}
\end{equation}
where $K$ denotes available knowledge. The quantity $P(I\mid K)$, the prior probability of the 
radiance $I$, may be absorbed into an overall normalization and does not concern us further.  It may 
be that the surface $T$ is of interest, whatever the value of emissivity.  In this case, one is free 
to denigrate $\epsilon_{k}$ as a "nuisance" parameter and integrate it out of (\ref{eqn3}), as will be 
done below.

Consider the remaining factors in (\ref{eqn3}) in turn, starting with the direct 
probability $P(I\mid T,\epsilon_{k} ,K)$ of observing radiance $I$ given $T$, $\epsilon_{k}$, 
and other \emph{a priori} knowledge $K$.  With aid of the forward model, it is possible to recast 
this quantity in more tractable form.  By hypothesis,
\begin{equation}
I(k)=I_{F}(k)+e_{k} \label{eqn4}
\end{equation}
where the error in spectral radiance $e_{k}$ is attributed to noise processes.  The prior probability 
for the noise $P(e_{k}\mid T ,\epsilon_{k} ,K)$ is now obtained by a maximum entropy argument.  If the 
noise power is assumed known, the noise probability is the function which maximizes the 
information-theoretic entropy subject to constraints imposed by the value of the noise power and 
overall normalization of probability,
\begin{eqnarray}
S=-\int_{-\infty}^{+\infty}P(e\mid K)log(P(e\mid K))de- \nonumber \\ 
\lambda_{1}\int_{-\infty}^{+\infty}e^{2}P(e \mid  K)de \label{eq:eqn5}
-\lambda_{2}\int_{-\infty}^{+\infty}P(e \mid K)de. 
\end{eqnarray}
The function maximizing (\ref{eq:eqn5}) is a Gaussian,
\begin{equation}
P(e\mid K)=\frac{1}{\sqrt{2\pi}\sigma}exp\left [-\frac{e^{2}}{2\sigma^{2}} \right] \label{eqn6}
\end{equation}
where the Lagrange multipliers for noise power and normalization have been written in terms of the RMS 
noise radiance $\sigma$.   Upon substituting (\ref{eqn4}) for the noise term, the probability of 
detecting a radiance $I$ given $T$, $\epsilon$, and noise $\sigma$ becomes
\begin{equation}
P(I\mid T,\epsilon,\sigma)=
exp\left[-\frac{(I-I_{F})^{2}}{2\sigma^{2}}\right]\frac{dI}{\sigma} \label{eqn7}
\end{equation}
This is also the likelihood function for $T$ and $\epsilon$. In order to formulate an estimator for 
$T$ and $\epsilon$, it remains to find the prior probability
\begin{equation}
P(T,\epsilon\mid K)=f(T,\epsilon)dTd\epsilon \label{eqn8}
\end{equation}
(The appropriate prior for noise is known to be the Jeffreys form, but is omitted here because it is 
assumed that the noise contribution is known.)  

We now follow Jaynes' prescription for finding an uninformative prior 
probability\cite{Jaynes1968},\cite{Jaynes1973},\cite{Jaynes1980}.  Assume two equivalent observers 
record the same sensor aperture radiance originating as thermal radiation from a surface, and 
interpret it in terms of Planckian emission characterized by a 
surface temperature and emissivities, subject to LOS attenuation.  Vladimir detects surface 
thermal emission $I$ in a solid angle $\Omega$, and describes the surface with parameters $T$ and 
$\epsilon$, and the attenuation with optical depth $\frac{\tau}{\mu}$: 
\begin{equation}
I=\epsilon_{k}B_{k}(T)exp(-\frac{\tau_{k}}{\mu}) \label{eqn9}
\end{equation}
He assigns prior probability in light of his knowledge regarding the problem
\begin{equation}
f(T,\epsilon)dTd\epsilon \label{eqn10}
\end{equation}
On the other hand, Estragon agrees with Vladimir on the definition of the Planck function, emissivity, and LOS 
attenuation, but describes the same situation with surface emission $I'$ in $\Omega$', and 
parameters $T'$, $\epsilon'$, and $\frac{\tau'}{\mu'}$, reporting
\begin{equation}
I'=\epsilon'_{k'}B_{k'}(T')exp(-\frac{\tau'_{k'}}{\mu'}) \label{eqn11}
\end{equation}
and assigning the prior probability
\begin{equation}
g(T',\epsilon')dT'd\epsilon' \label{eqn12}
\end{equation}
In order for the pair to agree as to the form of the estimator, the priors must be related by
\begin{equation}
g(T',\epsilon')dT'd\epsilon'=J^{-1}f(T,\epsilon)dTd\epsilon \label{eqn13}
\end{equation}
where
\begin{equation}
J=det\left[\frac{\partial{(T',\epsilon')}}{\partial{(T,\epsilon)}}\right] \label{eqn14}
\end{equation}
is the Jacobian determinant for the transformation between descriptions in the parameter space.

Assuming both are sober, Vladimir and Estragon must always be able to relate their descriptions of 
the sensed radiance by a Lorentz transformation.  Let us consider active transformations for 
concreteness.  Suppose that Vladimir wishes to describe events in a frame of reference moving at 
velocity $\beta=v/c$ along the observation axis, denoted $x$, with respect to the frame preferred by 
Estragon.  (It is convenient, although not actually necessary, to suppose that the $x$ axis is also 
the axis of photon propagation.)  Lorentz invariance requires that Vladimir's (unprimed) and Estragon's 
(primed) description of events be invariant under the Lorentz transformation given by
\begin{equation}
x'=\gamma(x-\beta ct) \label{eqn15}
\end{equation}
\begin{equation}
t'=\gamma(t-\beta x/c) \label{eqn16}
\end{equation}
where
\begin{equation}
\gamma=\frac{1}{\sqrt{1-\beta^{2}}}. \label{eqn17}
\end{equation}
The four-momentum of a photon travelling along the x-axis is
\begin{equation}
\mathsf{p}=  \left ( \begin{array}{c}
\hbar k/c \\	
\hbar k \\ \label{eqn18}
0 \\
0 \end{array} \right).
\end{equation}
Applying (\ref{eqn15}) and (\ref{eqn16}) to the components of (\ref{eqn18}), we see that Estragon and 
Vladimir relate their description of frequency or wavenumber by
\begin{eqnarray}
k'=\gamma (1-\beta) k \nonumber \\ 
\equiv \eta k. \label{eqn19}
\end{eqnarray}
 
How does the pair relate their descriptions of radiance?   Let a bundle of $\delta n$ photons with 
mean energy $p^{0}=\hbar ck$ and uncertainty $\delta p^{0}=\hbar c \delta k$ originate in a small area 
$\delta A$ of the radiating surface in a small time interval $\delta t$, collimated within a small 
solid angle $\delta \Omega$, and propagate unattenuated to an observer.  The surviving photons 
arriving at the observer's location comprise a collisionless photon gas.  A single photon in the 
bundle occupies a phase space volume 
\begin{equation}
V_{x}\times V_{k}=\delta A(c\delta t)\times\hbar^{3}k^{2}\delta k\delta\Omega\label{eqn20}
\end{equation}
while the bundle occupies a $6 \, \delta n$-dimensional phase space volume
\begin{equation}
V_{phase}=\left[ V_{x}V_{k}\right ]^{\delta n}. \label{eqn21}
\end{equation}
Equation (\ref{eqn21}) is invariant at any point on a photon trajectory.  According to a standard 
result in statistical physics, Liouville's 
theorem (\cite{LandauLifschitz1958}, pp. 9-10; 178), it has the same value at every point on that trajectory.  
So long as the photons remain 
collisionless they can neither leave their original volume of phase space, nor enter another.  
However Vladimir or Estragon choose to describe the patch of emitting surface $\delta A$, the time 
interval $\delta t$, the solid angle interval $\delta\Omega$ or the photon wavenumber $k$, they must 
agree as to the number of photons $\delta n$ in the bundle.  Hence, both (\ref{eqn20}) and the ratio
\begin{equation}
N=\frac{\delta n}{V_{x}V_{k}}=
\frac{\delta n}{\hbar ^{3}\delta A\delta tk^{2}\delta k\delta\Omega} \label{eqn22}
\end{equation} 
are invariant along a photon trajectory. 
 
Spectral radiance is defined as
\begin{equation}
I_{k}\equiv\frac{d(energy)}{d(time)d(area)d(frequency)d(solid angle)}. \label{eqn23}
\end{equation}
Rewriting the radiance as
\begin{equation}
I_{k}=\frac{\hbar k\delta n}{\delta A\delta t \delta k\delta \Omega}=\hbar^{4} k^{3} N, \label{eqn24}
\end{equation}
gives
\begin{equation}
\frac{I_{k}}{k^{3}}=\mbox{invariant} \label{eqn25}
\end{equation}
for any component of the total radiance along a given line of sight\cite{MTW1973}.  Equation
(\ref{eqn25}) has the same value in any 
frame of reference\cite{Weinberg1972}, with two consequences for this problem:

1.  The Planck function obeys
\begin{equation}
\frac{B_{\eta k}(\eta T)}{\eta^{3}}=B_{k}(T) \label{eqn26}
\end{equation}
as may also be seen by direct substitution in
\begin{equation}
B_{k}(T)=\frac{1}{\pi^{2}} \frac{k^{3}}{\left [exp\left [\frac{\hbar ck}{k_{B} T} \right]-1 \right ]}.
 \label{eqn27}
\end{equation} 
2.  Vladimir and Estragon must agree that the attenuated surface emission obeys
\begin{equation}
\frac{\epsilon'_{k'}B_{k'}(T')exp(-\frac{\tau'_{k'}}{\mu'})}{k'^{3}}=
\frac{\epsilon_{k}B_{k}(T)exp(-\frac{\tau_{k}}{\mu})}{k^{3}} \label{eqn28}
\end{equation}

Consider first the case of no attenuation, $\tau=0$.  Then
\begin{equation}
\frac{\epsilon'_{k'}B_{k'}(T')}{k'^{3}}=
\frac{\epsilon_{k}B_{k}(T)}{k^{3}} \label{eqn29}
\end{equation}
or
\begin{equation}
\epsilon'_{k'}B_{k'}(T')=\eta^{3}\epsilon_{k}B_{k}(T). \label{eqn30}
\end{equation}
One also has, from (\ref{eqn26}),
\begin{equation}
B_{\eta k}(T')=\eta^{3}B_{k}(T'/\eta). \label{eqn31}
\end{equation}

Combining (\ref{eqn19}), (\ref{eqn30}), and (\ref{eqn31}) gives
\begin{equation}
\epsilon'_{k'}B_{k}(T'/\eta)=\epsilon_{k}B_{k}(T). \label{eqn32}
\end{equation}
Now, Vladimir and Estragon also agree that, while they must lie between 0 and 1, emissivities are 
otherwise completely arbitrary functions of wavenumber, and by hypothesis have no dependence upon 
temperature.  In
\begin{equation}
\frac{B_{k}(T'/\eta)}{B_{k}(T)}=\frac{\epsilon_{k}}{\epsilon'_{\eta k}} \label{eqn33}
\end{equation}
the right-hand side can have no dependence upon $T$ or $T'$, while the left-hand side cannot be an 
arbitrary function of $\eta$ or $k$.  The only remaining possibility is
\begin{equation}
\frac{B_{k}(T'/\eta)}{B_{k}(T)}=\frac{\epsilon_{k}}{\epsilon'_{\eta k}}=\mbox{const.} \label{eqn34}
\end{equation}
The set of Lorentz transformations forms a group\cite{Einstein1905}, so this relation holds for 
the identity with $\beta=0$, $\gamma = \eta = 1$.  The constant must therefore equal unity.  

Next allow $\tau$ to differ from zero in (\ref{eqn28}).  In
\begin{equation}
\frac{k^{3}\epsilon'_{k'}B_{k'}(T')}{k'^{3}\epsilon_{k}B_{k}(T)}=
\frac{exp(-\frac{\tau_{k}}{\mu})}{exp(-\frac{\tau'_{k'}}{\mu'})} \label{eqn35}
\end{equation}
the left-hand side has, by hypothesis, no dependence upon LOS transmission, while the right-hand has 
no dependence upon surface properties so, again, both sides equal a constant, and, from 
(\ref{eqn26}) and (\ref{eqn34}), we find
\begin{equation}
\frac{k^{3}\epsilon'_{\eta k}B_{k'}(T')}{k'^{3}\epsilon_{k}B_{k}(T)}=
\frac{\epsilon'_{k'}B_{\eta k}(\eta T)}{\epsilon_{k}\eta^{3}B_{k}(T)}=1 \label{eqn36}
\end{equation}
or
\begin{equation}
\frac{exp(-\frac{\tau_{k}}{\mu})}{exp(-\frac{\tau'_{k'}}{\mu'})}=1 \label{eqn37}
\end{equation}
LOS attenuation does not affect the validity of (\ref{eqn34}).

Thus, the most general relation which respects a Lorentz transformation carrying wavenumber 
$k$ to $k'= \eta k$ is
\begin{equation}
T'=\eta T \label{eqn38}
\end{equation} 
\begin{equation}
\epsilon'_{k'}=\epsilon'_{\eta k}=\epsilon_{k}. \label{eqn39}
\end{equation}
The Jacobian is therefore
\begin{equation}
J=\eta \label{eqn40}
\end{equation} 
and
\begin{equation}
f(T,\epsilon_{k})=\eta g(\eta T,\epsilon'_{\eta k}). \label{eqn41}
\end{equation}

Invocation of the principle of indifference\footnote{As given by Jaynes\cite{Jaynes1968}, p. 128, in 
an extension of the original concept introduced by Laplace to encompass indifference between 
descriptions by distinct but equally cogent observers.} to assert Estragon and Vladimir must use the 
identical description of events, and thus assign the same prior probabilities,
\begin{equation}
f(T,\epsilon)=g(T,\epsilon) \label{eqn42}
\end{equation} 
leads to the functional equation
\begin{equation}
f(T,\epsilon_{k})=\eta f(\eta T,\epsilon_{k}). \label{eqn43}
\end{equation}
The solution of (\ref{eqn43}) is 
\begin{equation}
f(T,\epsilon)=\frac{\mbox{const.}}{T} \label{eqn44}
\end{equation}
yielding
\begin{equation}
f(T,\epsilon)dTd\epsilon=\frac{\mbox{const.}}{T}dTd\epsilon \label{eqn45}
\end{equation}
for the prior probability.

One now argues this form of the prior is least informative as to emissivity.  No functional dependence
upon a parameter should enter the form of the prior probability that is not imposed by the 
requirements of invariance and indifference.  Any such dependence would amount to the admission that 
we possess additional knowledge about emissivity beyond that assumed. That is, (\ref{eqn45}) is the 
unique choice of prior probability that assumes nothing about the value of $\epsilon_{k}$ beyond what 
is dictated by the problem statement.  A standard argument (found, for example, in 
\cite{Bretthorst1987}, pp. 9-15) then shows that prior knowledge about limits on the value of 
emissivity should appear in the limits of integration used in constructing marginal distributions for 
$T$. 

Surface temperature thus obeys the Jeffreys prior, while emissivity obeys the Bayes prior.  Both 
results may appear somewhat surprising, especially that for emissivity.  From the manner in which it 
appears in the expression for radiance, one's naive expectation might be that emissivity is a scale 
parameter.  However, the relation between the description of emissivity as seen by Vladimir and 
Estragon more resembles what one would expect of a location parameter: They must agree on the value 
of emissivity, but are free to assign it to different wavenumbers.

The result just obtained will now be extended to the situation in which radiance is sensed in bands 
wide enough that that it cannot be regarded as a function of wavenumber, but must be treated as an 
integral over a passband.  One then writes, for the contribution of surface emission to the total 
radiance at the sensor aperture in band i,
\begin{equation}
\int_{k_{1}}^{k_{2}} \!\!\!\epsilon_{k}B_{k}(T)exp(-\frac{\tau_{k}}{\mu})dk
\equiv\epsilon_{i}\int_{k_{1}}^{k_{2}} \!\!\!B_{k}(T)exp(-\frac{\tau_{k}}{\mu})dk \label{eqn46}
\end{equation}
It is always possible to do this by the mean value theorem for integrals, and it is frequently the 
case that the right-hand side of (\ref{eqn46}) expresses all available knowledge concerning the 
radiant properties of the emitting surface. 

Vladimir describes the surface emission by
\begin{equation}
\epsilon_{i}\int_{k_{1}}^{k_{2}}\!\!\!\ B_{k}(T)exp(-\frac{\tau_{k}}{\mu})dk=\int_{k_{1}}^{k_{2}} 
\!\!\!\epsilon_{k}B_{k}(T)exp(-\frac{\tau_{k}}{\mu})dk \label{eqn47}
\end{equation}
with

\vspace{3mm}
$
\;\;\;\;\;\;\;\;\;\;\;\;\epsilon_{k}=\left\{ \begin{array}{c}
0,k<k_{1} \\
\epsilon_{i},k_{1}\le k \le k_{2} \\
0,k>k_{2} 
\end{array} \right .
$
\vspace{3mm}

while, by (\ref{eqn37})-(\ref{eqn39}), Estragon describes things by
\begin{eqnarray}
\lefteqn{\int_{\gamma k_{1}}^{\gamma k_{2}}\epsilon'_{k'} 
B_{k'}(\gamma T)exp(-\frac{\tau'_{k'}}{\mu'})dk'} \nonumber \\ 
& & \equiv\epsilon_{i}'\int_{\gamma k_{1}}^{\gamma k_{2}} 
B_{k'}(\gamma T)exp(-\frac{\tau'_{k'}}{\mu'})dk' \label{eqn49}
\end{eqnarray}
with

\vspace{3mm}
$
\;\;\;\;\;\;\;\;\;\;\;\;\epsilon'_{k'}=\left \{ \begin{array}{c}
0,k'<\gamma k_{1} \\
\epsilon_{i},\gamma k_{1}\le k' \le \gamma k_{2} \\
0,k'>\gamma k_{2} 
\end{array} \right .
$
\vspace{3mm}

Comparison of the two expressions for surface emission, (\ref{eqn47}) and (\ref{eqn49}), leads to 
the immediate conclusion that the Jacobian connecting the two descriptions of surface temperature and 
band emissivity is
\begin{equation}
J=det\left[\frac{\partial{(T',\epsilon')}}{\partial{(T,\epsilon)}}\right] =\gamma, \label{eqn51}
\end{equation}
and
\begin{equation}
f(T,\epsilon)dTd\epsilon=\frac{\mbox{const.}}{T}dTd\epsilon \label{eqn52}
\end{equation}
once more.  

The result just obtained allows us to derive estimators for surface temperature and emissivity.  The 
starting point is a calculation of the marginal posterior probability for T given 
observed radiance in a finite number of bands when the surface emissivity in 
band $i$ is known to lie between $\epsilon_{min}(i)$ and  $\epsilon_{max}(i)$.  This quantity 
is computed for each band by integrating (\ref{eqn3}) between these limits, upon inserting 
(\ref{eqn7}) and 
(\ref{eqn45}).  Evaluating the integral requires completing the square in the exponent of 
(\ref{eqn7}).  To accomplish this, define auxilliary quantities $a$, $b$, and $c$, obtained from 
(\ref{eqn2}): 
\begin{equation}
a=\left [\int_{k_{1}}^{k_{2}}\left (B_{k}(T)-\frac{1}{\pi} F_{k}^\downarrow(0)
\right)exp(-\frac{\tau_{k}}{\mu})dk \right ]^{2}, \label{eqn53}
\end{equation}
\begin{equation}
 b=b_{1} b_{2} \label{eqn54}
\end{equation}
with
\begin{equation}
b_{1}=2\left [\int_{k_{1}}^{k_{2}}\left (B_{k}(T)-\frac{1}{\pi} F_{k}^\downarrow(0)
\right)exp(-\frac{\tau_{k}}{\mu})dk \right ] \label{eqn54a}
\end{equation}
\begin{equation}
b_{2}=\left[\int_{k_{1}}^{k_{2}}\left 
(\frac{1}{\pi} F_{k}^\downarrow(0)exp(-\frac{\tau_{k}}{\mu})+I_{k}^\uparrow(\tau,\mu) 
\right)dk-I_{i} \right ], \label{eqn54b}
\end{equation}
and
\begin{equation}
c=\left[\int_{k_{1}}^{k_{2}}\left 
(\frac{1}{\pi} F_{k}^\downarrow(0)exp(-\frac{\tau_{k}}{\mu})+
I_{k}^\uparrow(\tau,\mu) \right)dk-I_{i} \right ]^{2}
\label{eqn55}
\end{equation}
Then (dropping subscript i for the moment) (\ref{eqn3}) and (\ref{eqn7}) give
\begin{equation}
P(T,\epsilon \mid I,\sigma)\propto \frac{1}{\sqrt{2\pi}\sigma} exp\left[-\frac{(a\epsilon^{2}+
b\epsilon+c)}{2\sigma^{2}}\right] \label{eqn56}
\end{equation}
The marginal distribution obtained by integrating over the nuisance parameter $\epsilon$ is
\begin{equation}
P(T\mid I,\sigma)\propto \frac{1}{\sqrt{2\pi}\sigma} \int_{\epsilon_{min}}^{\epsilon_{max}} 
\!\!\!\exp\left[-\frac{(a\epsilon^{2}+b\epsilon+c)}{2\sigma^{2}}\right] d\epsilon \label{eqn57}
\end{equation}
Completing the square in the exponent allows this to be written as
\begin{equation}
P(T\mid I,\sigma)\propto \frac{1}{\sqrt{a}} exp \left [-\frac{\left [c-b^{2}/4a \right]}
{2\sigma^{2}} \right ] H(\epsilon_{max},\epsilon_{min}) \! \label{eqn58}
\end{equation}
where
\begin{eqnarray}
H(\epsilon_{max},\epsilon_{min})=
erf\left [\frac{\sqrt{a/2}(\epsilon_{max}+b/2a)}{\sigma} \right] \nonumber \\ 
-erf\left [\frac{\sqrt{a/2}(\epsilon_{min}+b/2a)}{\sigma} \right] \label{eqn59}
\end{eqnarray}
for each band i.  In (\ref{eqn59}) the error function is
\begin{equation}
erf(x)=\frac{2}{\sqrt{\pi}} \int_{0}^{x} exp \left (-t^{2} \right ) dt  \label{eqn59p5}
\end{equation}
  The joint posterior probability for observing radiances $I_{i}, i=1,n$ is
\begin{equation}
P(T \mid I_{i,i=1,n},\sigma)=\prod_{i=1}^{n} P(T \mid I_{i},\sigma). \label{eqn60}
\end{equation}
Assuming $T$ is known to lie between a minimum and a maximum, an estimator for T given radiance in 
band i is
\begin{equation}
\langle T \rangle = \frac{\int_{T_{min}}^{T_{max}} TP(T \mid I_{i},\sigma)\frac{dT}{T}}
{\int_{T_{min}}^{T_{max}} P(T \mid I_{i},\sigma)\frac{dT}{T}} \label{eqn61}
\end{equation}
while a joint estimator for T given radiances in all n bands is
\begin{equation}
\langle T \rangle = \frac{\int_{T_{min}}^{T_{max}}TP(T \mid I_{i,i=1,n},\sigma)\frac{dT}{T}}
{\int_{T_{min}}^{T_{max}}P(T \mid I_{i,i=1,n},\sigma)\frac{dT}{T}} \label{eqn62}
\end{equation}  
An estimator for the emissivity in band $i$ is given by
\begin{equation}
\langle \epsilon_{i} \rangle =
 \frac{\int_{\epsilon_{min}}^{\epsilon_{max}}\epsilon P(\langle T \rangle ,\epsilon
 \mid I_{i},\sigma)d\epsilon}
{\int_{\epsilon_{min}}^{\epsilon_{max}}P(\langle T \rangle ,\epsilon \mid I_{i},\sigma)d\epsilon} 
\label{eqn63}
\end{equation}
This form has the advantage that estimates of the surface $T$ are significantly less sensitive to 
discrepancies between sensed and modeled radiances than are estimates of emissivity.  An estimate of 
$T$ obtained from (\ref{eqn58})-(\ref{eqn62}) with 
uninformative limits on emissivity may be close enough in practice for accurate emissivity retrievals 
by (\ref{eqn63}).  (Equation (\ref{eqn63}) may be evaluated in closed form with elementary functions; 
however, the resulting expression is quite cumbersome and is omitted here.)

\section{Monte Carlo Simulation of MODIS Land Surface Temperature Retrieval}
A land surface temperature retrieval algorithm has been developed using the results just derived.
While the intent of the work reported in this paper is to unshackle LST estimation from emissivity 
knowledge, the algorithm also retrieves emissivity estimates, and may be thought of as a TES algorithm
if desired.  It is intended to illustrate the application of Bayesian analysis to thermal remote 
sensing, and is 
purported to be optimal neither in execution speed nor accuracy.  The 
requirement for only one forward model calculation per retrieval suggests that it will not impose an 
extreme computational burden in practice (even though the forward model to be 
described requires two MODTRAN calculations).  The algorithm is used to simulate LST retrieval from a 
notional exoatmospheric sensor that records radiance from a patch of the Earth's surface at a 
specified signal-to-noise ratio(SNR).  It is assumed that the dominant noise contribution arises from 
the shot noise of the radiance incident upon the sensor aperture.   

In outline, the algorithm works as follows.  Equation (\ref{eqn58}) gives the distribution of surface 
temperature consistent with observed radiances and the initial range of emissivities.  This 
distribution typically differs from zero only within a narrow range about the true surface 
temperature.  Within this range, (\ref{eqn62}) is used for each of $n$ bands and for all 
bands jointly to compute $n+1$ separate estimates of the surface temperature.  The actual surface
temperature  is assumed to lie between the extreme values of this set of expected values, which now 
determine the allowable range.  The joint temperature distribution and the various expectation values 
for surface temperature are next refined using the contracted range of \emph{a-priori} credible surface 
temperatures in (\ref{eqn58}), now calculated with a finer temperature mesh.  After a few iterations 
of this procedure, the different surface temperature expectation values obtained from (\ref{eqn61}) and
 (\ref{eqn62})
reliably converge to a single value lying close to the true surface temperature.  A convergence  
radius $\eta$
for the different estimates of 0.01 K was used.  Emissivities are then obtained by substitution of the 
joint surface temperature estimate into the expression for band emissivity expectation values, 
(\ref{eqn63}).  
Equation (\ref{eqn56}) being a Gaussian distribution, it is possible to refine the 
\emph{a-priori} limits on credible emissivites by specifying a threshold of $m$ standard deviations.  
Six is used for the examples presented here.  The revised \emph{a-priori} emissivity limits and 
surface temperature limits may then be used to restart the entire sequence just outlined, if 
desired.  This additional iterative loop was repeated once in the simulation presented here.  Surface 
temperature and emissivity values show only marginal changes as a result of the second iteration, indicating 
convergence of the retrieval.
	
The starting point is the posterior distribution for surface temperature (\ref{eqn58}).  To compute 
it, the coefficients $a(T)$, $b(T)$, and $c$  are obtained as a 
function of surface temperature.  Isaacs two-stream MODTRAN4 calculations with SALB=1.0 supply those 
forward-model spectral quantities independent of surface temperature:  Attenuation along the 
line-of-sight, downwelling radiance at the surface, and upwelling radiance at TOA.  The surface 
thermal emission component is computed directly from the Planck function and spectral attenuation.  A 
further MODTRAN calculation with SALB=0.0 is used to obtain estimates of scattered solar radiance and
 of the "scattered thermal" radiance compoment of the total radiance returned by MODTRAN.  

The calculation of $a(T)$ and $b(T)$ used in the retrievals departs from (\ref{eqn53})-(\ref{eqn54a}) 
in one regard.  Approximation of the total TOA radiance computed by MODTRAN with the (never exact) 
Duntley equation becomes increasingly inaccurate as the surface temperature increases.  The dominant 
contribution to the discrepancy is the surface emission portion of the scattered thermal radiance.  
Subtracting an estimate of this term gives a corrected TOA radiance in much better accord with the 
predictions of the Duntley equation.  The estimate is calculated as a function of the unknown surface 
emissivity and (necessarily erroneous) forward model boundary temperature, using the 
difference between 
the scattered thermal contributions computed with SALB=1.0 and SALB=0.0.  Rather than subtracting the 
approximate scattered surface thermal radiance from the total MODTRAN radiance, the correction was 
implemented in a mathematically equivalent way by adding that portion of the scattered thermal 
radiance component linear in surface emissivity to the surface thermal emission terms in 
(\ref{eqn53})-(\ref{eqn54a}).

It should be noted that while the forward model assumes knowledge 
of atmospheric parameters, MODTRAN-computed quantities used in the forward model have no dependence 
upon true surface temperature or emissivity.  The boundary 
temperature parameter used in MODTRAN affects the forward radiance calculations primarily through the
scattered thermal contribution.  In addition, MODTRAN adjusts the 
atmospheric temperature profile in the lower zones to interpolate smoothly between surface conditions 
and a fiducial layer in the atmosphere.  For these reasons the forward model boundary temperature 
should not differ greatly from a physically reasonable value.

The algorithm is executed according to these steps:

1.  Perform the forward model radiative transfer calculations with MODTRAN.

2.  Calculate the individual band posterior probabilities, and the joint posterior probability over
all bands, as a function of surface temperature with (\ref{eqn61}) and (\ref{eqn62}), contracting as
necessary the range of T to that giving nonvanishing joint posterior probability in (\ref{eqn60}).  

3.  Calculate expectation values for surface temperature over the posterior probabiltiy for each band
individually, and over the joint posterior probabiltity.  This calculation gives $n+1$ surface 
temperature estimates.

4.  Perform convergence test $max |\langle T_{i} \rangle -\langle T_{j} \rangle| < \eta$ over all 
pairs of surface temperature estimates.  If the convergence test is satisfied, proceed to step 5.  
Otherwise, iterate by repeating steps 2-4.

5.  Compute expectation values for band emissivities using (\ref{eqn63}).

6.  Adjust $\epsilon_{min}, \epsilon_{max}$ to $\pm m$ standard deviations about 
$\langle \epsilon_{i} \rangle$ for each band.

(7.  Repeat steps 2-6 if desired.)

Monte Carlo simulations of LST retrieval in a selected subset of MODIS bands illustrate the 
performance of the algorithm.  The bands chosen appear in Table 1.  MODTRAN calculations are used 
both as simulated TOA radiances in MODIS bands and as the forward model.  Each Monte Carlo realization 
of TOA radiance is calculated using a mid-latitude summer atmosphere with MODTRAN parameters listed in 
Table 2 selected as uniform deviates 
within the limits shown, including "true" surface T and band emissivities.  It is unlikely that the 
atmospheric profile or other parameters required to
specify a formard radiative transfer model will be reliably known to high accuracy.  In order to 
simulate the effect of imperfect knowledge on the forward model, a second draw of random numbers is 
used to  introduce errors in the fallible forward model as shown in 
the last column.  Thus, for examples, the surface visibility, which MODTRAN uses to parameterize 
aerosol effects, is chosen to lie between 5 and 30 km for each Monte Carlo realization.  A random error 
of up to $\pm 4$ km is added to this value of visibility for use in calculation of the fallible forward 
model for each realization.  The MODTRAN model default water vapor profile, given as (grams 
precipitable water)/ (kilograms air), 
is randomly scaled between the limits shown (subject to the constraint that relative humidity cannot 
exceed $100 \%$).  The range of perturbed forward model parameters is 
truncated at the limits 
specified in Table 2.  The forward model cannot, of course, incorporate knowledge of the true surface 
temperature, but it does require an initial guess for that quantity.  This guess is obtained by 
varying the forward model boundary temperaature randomly from "truth" by $\pm 20 K$, 
without truncation.  Inclusion of the scattered thermal radiance contribution in the forward model 
notably improves retrieval accuracy, despite boundary temperature uncertaintiies of this magnitude in
the forward model.

\begin{table}[h]
Table 1.  MODIS bands used in simulations 
{\small \begin{tabbing}
MODIS band...\=.......wavelength/limits\=....notional/snr  \kill
MODIS band\>wavelength limits\>notional snr \\ 
\rule{70mm}{0.1mm} \\
    20 \> 3.660-3.840 $\mu$ \> 350 \\
22\> 3.929-3.989 $\mu$ \> 350 \\
23\> 4.020-4.080 $\mu$ \> 350  \\
29\> 8.400-8.700 $\mu$ \> 1000 \\
31\> 10.870-11.280 $\mu$ \> 1000 \\
32\> 11.770-12.270 $\mu$ \> 1000  \\
\end{tabbing}}
\end{table} 

\begin{table}[h]
Table 2.  Monte Carlo parameter ranges 
{ \small \begin{tabbing}
MODTRAN parameter\=..\=.....min./value\=.max./value    
\=....forward/model/error  \kill
MODTRAN \>\>minimum \>maximum\>fallible forward \\
parameter\>\>value\>value\>model error \\ 
\rule{87mm}{0.1mm} \\
nadir view angle\>\>125 $\deg$\>180 $\deg$\>$\pm 0.125 \, \deg$ \\
surface visibility\>\>5 km \>30 km \> $\pm$ 4 km \\
column water vapor\>\>0.33 $\times$ MLS\>1.00\> $\pm 0.2$  \\
thin cirrus altitude\>\>8 km \>12 km \>$\pm 0.5$ km \\
thin cirrus thickness\>\>1 m \>20 m\> $\pm 25$ m \\
thin cirrus opacity\>\>$0.05/km$\>$0.2/km$\> $\pm 0.025/km$  \\
solar azimuth\>\>0 $\deg$\>90 $\deg$\> $\pm 0.125 \, \deg$  \\
solar elevation\>\>20 $\deg$\>60 $\deg$\> $\pm 0.125 \, \deg$  \\
viewng azimuth\>\>0 $\deg$\>90 $\deg$\> $\pm 0.125 \: \deg$  \\
viewing elevation\>\>35 $\deg$\>90 $\deg$\> $\pm 0.125 \, \deg$  \\
surface T\>\>268 $\deg$ K\>328 $\deg$ K\> N/A \\
\end{tabbing}}
\end{table}

Next, the simulated MODIS TOA radiances are contaminated with notional sensor noise simulated as a
zero-mean Gaussian random process with standard deviation equal to the noise equivalent radiance 
$NE\Delta R$.  The algorithm also requires an estimated variance for the noise radiance 
($\sigma$ in (\ref{eqn6})), which should be of order $NE\Delta R$.  $NE\Delta R$ is 
parameterized in terms of a 
signal-to-noise ratio.  SNR was chosen to lie on the low end of values inferred from MODIS 
$NE\Delta T$ values\cite{Guenther2002} with the aid of the following estimate.  The error in TOA 
radiance from noise sources is estimated as, roughly, 
\begin{eqnarray}
\delta I =\frac{\partial{I}}{\partial{T}}\delta T+O(\delta T)^{2} \\ \nonumber
\cong \frac{\partial{I_{s}}}{\partial{T}} NE\Delta T \label{eqn66}
\end{eqnarray}
where
\begin{equation}
I_{s}=\epsilon \int_{\Delta k}B_{k}(T)exp(-\frac{\tau_{k}}{\mu})dk \label{eqn67}
\end{equation}
is the attenuated surface thermal emission at TOA, leading to 
\begin{equation}
SNR=\left (\frac{\delta I}{I} \right)^{-1} \cong \left 
(\frac{\partial{log(I_{s})}}{\partial{T}} NE\Delta T \right )^{-1} \label{eqn68}
\end{equation}
This quantity was computed for each Monte Carlo realization; the SNR values used for the retrievals, 
which apppear in Table 1, conservatively underestimate (\ref{eqn68}).  Performance of the algorithm 
appears not too sensitive to the exact noise contamination added to the simulated band radiances, nor 
to the exact noise variance assumed in the retrieval, as long as neither is grossly erroneous.  In 
fact, it is 
possible to adjust the assumed value of $\sigma$ in the estimator to contract or expand the range of 
viable surface 
temperatures consistent with sensor radiances without disastrously biasing the retrieval, as described 
below.

One thousand Monte Carlo realizations each were calculated for day and night, with a mid-latitude 
summer atmosphere.  Generous bounds for the initial \emph{a-priori} limits on LST and band emissivity 
were assumed, subject to the physical upper bound on surface emissivity:
\begin{equation}
\begin{array}{c}
200 K \le T \le 500 K \\ \nonumber
0.75 \le \epsilon_{i} \le 0.99
\end{array} \label{eqn69}
\end{equation}
Note that  both limits for LST lie outside the range sampled by 
the Monte Carlo draws.  

\begin{table}[h]
Table 3.  Monte Carlo simulation results: mean errors, \\
standard deviations 
{ \small \begin{tabbing}
Case\=................\=.................Day....\=...................Night    \kill
Case\>\>Day\>Night \\
\rule{65mm}{0.1mm} \\
LST error(K)\>\>$-0.25 \pm 1.23$ \>$-0.31 \pm 1.11$ \\
$\epsilon_{20}$ error\>\>$-0.004\pm 0.022$ \>$-0.003\pm 0.035$ \\
$\epsilon_{22}$ error\>\>$-0.009\pm 0.034$ \>$+0.001\pm 0.034$ \\
$\epsilon_{23}$ error\>\>$-0.008\pm 0.048$ \>$-0.007\pm 0.038$  \\
$\epsilon_{29}$ error\>\>$-0.004\pm 0.031$ \>$-0.003\pm 0.022$ \\
$\epsilon_{31}$ error\>\>$-0.005\pm 0.023$ \>$-0.005\pm 0.022$ \\
$\epsilon_{32}$ error\>\>$-0.007\pm 0.028$ \>$-0.006\pm 0.029$  \\
\end{tabbing}}
\end{table} 

Mean errors and error 
standard deviations for the retrieved surface temperatures and band emissivities, with respect to 
"true" values, appear in Table 3.
In the majority of cases, acceptable estimates of LST and band emissvities were obtained using all 
six bands from Table 2, with the SNR chosen to equal the assumed noise variance in the simulations 
which appears in Table 1.  However, in about $4\%$ of the simulations it proved impossible to find 
an acceptable solution with all six bands in this manner.  The solution instead tended to badly 
erroneous values (e.g., $T \le 100 K$, and emissivities pegged at the limits of the 
prior).  Inspection of the posterior probabilities from (\ref{eqn61}) revealed that in these anomalous 
cases one or more of the individual band posterior probabilities fails to overlap significantly with 
the product of the remaining posterior distributions, leading to a joint posterior probability that 
effectively vanishes.  A number of remedies is available when this difficulty arises.  The number of 
successful retrievals rises sharply when the range of the band emissivity  prior is expanded to 
0.7-0.999, but at the cost of reducing their overall accuracy somewhat.  Experimentation 
shows that, in all of the anomalous cases, it is possible to get a satisfactory LST retrieval with 
some three-band subset of the original six.  The LST so obtained can then be inserted into the 
expectation value for band emissivity to yield emisssivities for all six bands.  Finally, the support 
of the joint posterior probability can be broadened by increasing the noise radiance $\sigma$ 
in its calculation.  This 
last approach was used to obtain Table 3.  The effect on retrievals of increasing $\sigma$ for 
subsets of the bands  
differed negligibly from that of increasing $\sigma$ for all bands by the same factor.  The most 
intractable
of the anomalous cases (one each nighttime and daytime) required increasing $\sigma$ by a factor 
of 7.0 
in order to obtain an acceptable solution.

Examination of the $\chi^{2}$ statistic for retrieval errors shows that the estimator is 
(slightly) biased for normal, and significantly biased for anomalous, retrievals.  For the 963 normal 
nighttime retrievals the mean and stardard deviation of the surface temperature error are 
$\delta T=-0.26 \pm 1.06 K$, with $\chi^{2}=$ 1.06 per degree 
of freedom.  For 958 normal daytime retrievals, the corresponding figures are 
$\delta T=-0.21 \pm 1.18 K$ and  $\chi^{2}=$1.03 per degree of freedom.  The values for 37 anomalous 
nighttime retrievals are $\delta T=-1.71 \pm 1.32 K$ and $\chi^{2}=$ 2.64 per degree of freedom, with 
$\delta T=-1.08 \pm 1.92 K$ and $\chi^{2}=$1.29 per degree of freedom, for 42 anomalous daytime cases.  However, the 
distribution of errors is very accurately Gaussian (as should be expected).  If the mean error is 
subtracted before calculating $\chi^{2}$, the result is identically 0.999 per degree of 
freedom for surface temperature (and all six band emissivities) for all retrievals.

Forward models for the anomalous cases systematically have large errors in one or more of the 
randomly-varied simulation parameters.  Thus, the surface visibility and the boundary 
temperature are both more likely to lie near the limits of their range than in the middle.  In 
particular, the column water vapor scaling factor is over three times as likely to exceed 1.1 for 
anomalous retrievals as for normal ones (25/37 vs 203/963 nighttime; 27/42 vs 182/958 daytime), with 
only three daytime and two nighttime anomalous retrievals occurring for a column water vapor scaling
less than unity.  The frequency of anomalous retrievals appears, to some extent, to be an artifact of 
inserting large errors in the forward model to approximately simulate imperfect knowledge of 
atmospheric conditions.  

In any event, all 2000 Monte Carlo realizations led to a successsful retrieval of both LST and six 
band emissivities. 

\section{Discussion}
Points which should be addressed in further developments of practical algorithms:

1. It appears that this approach to TES works largely because the range of plausible surface 
temperature values consistent with band radiances and an uninformative range of band emissivity is 
quite constricted, as a consequence of the strong temperature dependence of the Planck function.  
It turns out not to be terribly difficult to get a temperature estimate that is close enough to truth 
that it can be inserted into the least sophisticated imaginable estimator for band emissivity 
(\ref{eqn63}), and still lead to acceptably accurate results.

Once the algorithm has gotten to an iteration in which the current range of temperature 
and band emissivities is restricted to a neighborhood sufficiently close to the true values that the 
posterior distribution is jointly 
Gaussian in $T$ as well as in the $\epsilon_{i}$, it is apparent both that convergence to the true 
values will occur as assumed in this algorithm, and that these values will maximize the likelihood.  
However, at present, there is no proof in hand that the procedure outlined above actually converges, 
or that, given that it does, it converges to the true surface temperature and emissivity 
combination.  It appears to do both to good accuracy in practice.

However, the algorithm did-initially-fail to converge to an acceptable solution in about $4\%$ of the 
realizations.  As recounted in the previous section, it proved possible in every case to adapt the 
search strategy so as to successfully retrieve both LST and emissivities for all bands.  The 
successful recovery strategies all had the effect of maximizing the numerical joint posterior 
probability, by some combination of 1) eliminating from the estimator band posterior probabilities 
whose effective nonvanishing support did not intersect that of the joint probability of the remaining 
bands, or forcing intersection by broadening the support of the outlier posterior probabilities by 
2) loosening limits on the prior, or 3) increasing the noise radiance parameter assumed in the retrieval.

The solutions obtained are, in any event, not unique, because the TES problem is underdetermined.  
Considered as a surrogate for maximum likelihood solutions, the algorithm solutions approximate only 
local maxima, and it might be possible to find maxima which give very poor account of temperature and 
emissivity.  This has not happened in simulations performed to date.

2.  The algorithm as presently formulated appears to be unnecessarily complicated.  It seems certain 
that its operation can be significantly streamlined.  For practical applications, it will be necessary 
to eliminate redundant elements of the calculation.

3.  The model for band radiances in this memo treated them independently, apart from the prior 
knowledge that the surface temperature for all bands must be the same.  Bretthorst 
\cite{Bretthorst1987},\cite{Bretthorst1988},\cite{Bretthorst1988a} has addressed problems involving more 
sophisticated models for 
observations, in an approach which would appear to offer real advantages in the present context.

4.  Perhaps the least satisfactory feature of this algorithmic approach is its dependence upon an 
accurate forward radiance model.  To the extent MODTRAN can be regarded as supplying radiance 
estimates which are zero-mean error estimates of the true radiance, the effect of radiance prediction 
error on this algorithm may simply be regarded as a contribution to the noise variance.  But in real 
life, a forward model can be expected to have systematic errors that need not originate as unbiased 
stationary Gaussian processes.  The question whch has been addressed in this work 
is:  Given an accurate forward model (in the sense just described), what surface temperature and band 
emissivities are consistent with observed radiances and knowledge of their error statistics?  A harder 
question, which will be the focus of further developments, is:  Given a fallible, but reasonably 
accurate, forward model, what surface temperatures and band emissivities can possibly be consistent 
with observations and available knowledge, no matter what the forward model error, so long as it falls 
within known limits?

\section{Conclusion}
A simple argument, based on inherent physical symmetries that the description of surface thermal 
emission must obey, leads to the appropriate prior probabilities for surface temperature and 
emissivity.  These lead to the maximum entropy estimator for the mismatch between sensed and
modeled radiance in the 
presence of noise, from which an estimator of surface temperature may be constructed that treats 
emissivity as a nuisance parameter.  MODTRAN-based simulations show that temperature-emissivity 
separation is successfully performed by iteration between the temperature estimator, and a similar 
estimator for surface emissivity.


\nocite{*}
\bibliographystyle{IEEE}

\begin{thebibliography}{1}
\bibitem{Dash2002}Dash, P., F.-M. G\"{o}ttsche, F.-S. Olesen, and H. Fischer, "Land surface 
temperature and emissivity estimation from passive sensor data:  theory and practice-current
trends," \emph{Int. J. Remote Sensing}, vol. 23, pp. 2563-2594, 2002

\bibitem{Wan1999}Wan, Z.-M., \emph{MODIS Land-Surface Temperature Algorithm Theoretical Basis 
Document}, Institute for Computational Earth System Science, University of California, 
Santa Barbara, 1999

\bibitem{Li1999}Li Z.-L.,F. Becker, M. P. Stoll, and Z. Wan, "Evaluation of Six Methods for Extracting 
Relative Emissivity Spectra from Thermal Infrared Images," \emph{Rem. Sens. Env.}, vol. 69, 
pp. 197-214, 1999

\bibitem{KahleAlley1992}Kahle, A. B., and R. E. Alley, "Separation of Temperature and Emittance in 
Remotely Sensed Radiance Measurements," \emph{Rem. Sens. Env.}, vol. 42, pp. 107-111, 1992

\bibitem{KealyHook1993}Kealy, P. S., and S. J. Hook, "Separating Temperature and Emissivity in
Thermal Infrared Multispectral Scanner Data:  Implications for Recovering Land Surface Temperatures," 
\emph{IEEE Trans. Geosci. Remote Sensing}, 
vol. 31, pp. 1155-1164, 1993

\bibitem{PetitcolinVermote2002}Petitcolin, F., and E. F. Vermote, "Land Surface
Reflectance, Emisivity and Temperature from MODIS Middle and
Thermal Infrared data," \emph{Rem. Sens. Env.},vol 83(1-2), 112-134, 2002  

\bibitem{LiBecker1993}Li Z.-L., and F. Becker, "Feasibility of Land Surface Temperature and Emissivity
Determination from AVHRR Data," \emph{Rem. Sens. Env.}, vol. 43, pp. 67-85, 1993

\bibitem{Watson1992}Watson, K., "Spectral Ratio Method for Measuring Emissivity," 
\emph{Rem. Sems. Env.}, vol. 42, pp. 113-116, 1992

\bibitem{BorelSzymanski1998}Borel, C. C., and J. . Szymanski, "Physics-based Water and Land 
Temperature Retrieval," in Handbook of Science Algorithms for the Multispectral Thermal Imager, B. W. 
Smith, Ed., Los Alamos National Laboratory and Savannah River Technology Center, 1998

\bibitem{WanDozier1996}Wan, Z.-M., and J. Dozier, "A generalized split-window algorithm for retrieveing 
land-surface temperature from space," \emph{IEEE Trans. Geosci. Remote Sensing}, 
vol. 34, pp. 892-905, 1996

\bibitem{BarducciPippi1996}Barducci, A., and I. Pippi, "Temperature and emissivity retrieval from
remotely sensed images using the 'Grey body emissivity' method," \emph{IEEE Trans. Geosci. Remote Sensing}, 
vol. 34, pp. 681-695, 1996

\bibitem{WanLi1997}Wan, Z.-M., and Z.-L. Li, "A physics-based algorithm for land-surface emissivity
and temperature from EOS/MODIS data," \emph{IEEE Trans. Geosci. Remote Sensing}, 
vol. 35, pp. 980-996, 1997

\bibitem{Duntley1948}Duntley, S. Q., "The Reduction of Apparent Contrast by the Atmosphere,"
\emph{J. Opt. Soc. Am} vol. 38, p 179, 1948

\bibitem{footnote1}It happens that Lambertian behavior is usually considered a 
good approximation in the LWIR.  A Lambertian assumption is regarded as more questionable in the MWIR.  
Looking ahead to later simulated MODIS retrievals, Wan and Li~\cite{WanLi1997} have examined this 
question for MODIS MWIR surface imaging bands (Bands 20,22, and 23) and have 
concluded that in MWIR bands surfaces may be adequately approximated as Lambertian reflectors obeying 
Khirchoff's law.  For the present application, the relevant point is the applicability of Khirchoff's 
law, rather than Lambert's. 

\bibitem{Bretthorst1987}Bretthorst, L., \emph{Bayesian Spectral Analysis and Parameter 
Estimation}, Dissertation, Washington University, St. Louis, MO, 1987

\bibitem{Bretthorst1988}Bretthorst, L., "Bayesian Spectrum Analysis and Parameter 
Estimation," in Berger, J., S. Fienberg, J. Gani, K. Krickenberg, and B. Singer, Eds, 
\emph{Lecture Notes in Statistics,} {\bf{48}}, Springer-Verlag, New York, 1988

\bibitem{Bretthorst1988a}Bretthorst, L., "Excerpts from Bayesian Spectrum Analysis and Parameter 
Estimation," in Erickson, G. J., and C. R. Smith, \emph{Maximumum-Entropy and Bayesian Methods in
Science and Engineering, Volume 1:  Foundations}, Kluwer, Dordrecht, 1988, pp. 75-145

\bibitem{LandauLifschitz1958}Landau, L. D., annd L. Lifschitz, \emph{Statistical Physics}, 
Addison-Wesley: Reading, MA, 1958

\bibitem{Jaynes1968}Jaynes, E., "Prior Probabilities," 
\emph{IEEE Trans. on Systems Science and Cybernetics}, vol. SSC-4, pp. 227-241, 1968

\bibitem{Jaynes1973}Jaynes, E., "The Well-Posed Problem," \emph{Found. Physics 3}, pp. 477-493, 1973

\bibitem{Jaynes1980}Jaynes, E., "Marginalization and Prior Probabilities," in \emph{Bayesian
Analysis in Econometrics and Statitstics}, A. Zellner, Ed., North-Holland Publishing Co.: 
Amsterdam, 1980

\bibitem{MTW1973}Misner, C. W., K. S. Thorne, and J. A. Wheeler, \emph{Gravitation}, Freeman: 
San Francisco, 1973

\bibitem{Weinberg1972}Weinberg, S., \emph{Gravitation and Cosmology}, John Wiley and Sons: 
New York, 1972

\bibitem{Einstein1905}Einstein, A., "Zur elektrodynamik bewegter K\"{o}rper," \emph{Ann. Physik},
vol. 17, pp. 891-921, 1905

\bibitem{Guenther2002}Guenther, B., X. Xiong, V. V. Salmonson, W. L. Barnes, and J. Young, 
"On-orbit perfromance of the Earth Observing System Moderate Resolution Spectroradiometer; 
first year of data," \emph{Rem. Sens. Env. 83}, pp. 16-30, 2002

\end{thebibliography}

\end{document}